\documentclass[reprint,superscriptaddress,prb,aps]{revtex4-2}

\usepackage{mathtools}
\usepackage{amsmath}
\usepackage{amsthm}
\usepackage{amssymb}
\usepackage{graphicx}
\usepackage{bm}
\usepackage{siunitx}
\usepackage[colorlinks,linkcolor=blue, urlcolor=blue,citecolor=blue,bookmarks=false]{hyperref}

\begin{document}

\title{Coexisting electronic smectic liquid crystal and superconductivity \\ in a Si square-net semimetal}

\author{Christopher J. Butler}
\email{christopher.butler@riken.jp}
\affiliation{RIKEN Center for Emergent Matter Science, 2-1 Hirosawa, Wako, Saitama 351-0198, Japan}

\author{Toshiya Ikenobe}
\affiliation{Institute for Solid State Physics, The University of Tokyo, Kashiwa, Chiba 277-8581, Japan}

\author{Ming-Chun Jiang}
\affiliation{RIKEN Center for Emergent Matter Science, 2-1 Hirosawa, Wako, Saitama 351-0198, Japan}
\affiliation{Department of Physics and Center for Theoretical Physics, National Taiwan University, Taipei 10617, Taiwan}

\author{Daigorou Hirai}
\affiliation{Department of Applied Physics, Nagoya University, Nagoya, Aichi 464-8603, Japan}

\author{Takahiro Yamada}
\affiliation{Institute of Multidisciplinary Research for Advanced Materials, Tohoku University, Sendai, Miyagi 980-8577, Japan}

\author{Guang-Yu Guo}
\affiliation{Department of Physics and Center for Theoretical Physics, National Taiwan University, Taipei 10617, Taiwan}
\affiliation{Physics Division, National Center for Theoretical Sciences, Taipei 10617, Taiwan}

\author{Ryotaro Arita}
\affiliation{RIKEN Center for Emergent Matter Science, 2-1 Hirosawa, Wako, Saitama 351-0198, Japan}
\affiliation{Department of Physics, The University of Tokyo, 7-3-1 Hongo Bunkyo-ku, Tokyo 113-0033, Japan}

\author{Tetsuo Hanaguri}
\email{hanaguri@riken.jp}
\affiliation{RIKEN Center for Emergent Matter Science, 2-1 Hirosawa, Wako, Saitama 351-0198, Japan}

\author{Zenji Hiroi}
\affiliation{Institute for Solid State Physics, The University of Tokyo, Kashiwa, Chiba 277-8581, Japan}

\begin{abstract}

Electronic nematic and smectic liquid crystals are spontaneous symmetry-breaking phases that are seen to precede or coexist with enigmatic unconventional superconducting states in multiple classes of materials. In this Letter we describe scanning tunneling microscopy observations of a short ranged charge stripe (smectic) order in NaAlSi, whose superconductivity is speculated to have an unconventional origin. As well as this we resolve a clear spatial modulation of the superconducting gap amplitude, which arises due to the intertwined superconducting and smectic orders. Numerical calculations help to understand the possible driving mechanism as a suppression of kinetic energy on the Fermi surface formed in part by two large, flat-topped hole pockets of $p$-orbital character.

\end{abstract}

\maketitle

Fluids of correlated electrons can spontaneously adopt liquid crystal-like distributions that break symmetries of the host crystal lattice. By analogy to ordinary liquid crystals, those that break the lattice's rotational symmetry are said to be nematic, while those that break both its rotational and translational symmetries are said to be smectic. A nematic electronic fluid was first proposed as a precursor to the superconducting state of a doped Mott insulator \cite{Kivelson1998}, and electronic liquid crystals more generally are thought to have a strong relationship, either competing or cooperative, with superconductivity (SC). They are observed to form within the $d$ bands of several classes of cuprate and Fe-based high-$T_{c}$ superconductors \cite{Tranquada1995,Lawler2010,Keimer2015,Shibauchi2020}. In the cuprates, scanning tunneling microscopy (STM) imaging famously shows that liquid crystal-like phases take on distinctive `checkerboard' or short-ranged ladder-like configurations \cite{Hanaguri2004,Kohsaka2007}. In other $d$-orbital systems they are seen to adopt more-or-less uniaxial density-of-states modulations or charge stripe configurations \cite{Chuang2010,Yim2018,Yuan2021,Butler2022,Li2023,Naritsuka2023,Wang2025}.

Similar liquid crystal behavior in $s$- or $p$-orbital systems, which do not typically exhibit strong correlations, is not generally anticipated. Nevertheless, recent investigations of Sb-based square-net semimetals have shown anomalous charge ordered phases, other than charge density waves (CDWs), that can closely resemble the ladder-like ordered patterns seen in the cuprates despite their purer $p$-orbital character \cite{Venkatesan2025,Que2025}. The reason for apparently similar ordered phases in disparate $p$- and $d$-orbital systems and, for the $p$ case in general, any potential relationship between such ordered phases and SC, remain to be understood.

Here we use STM to examine NaAlSi, a nodal-line semimetal \cite{Muechler2019,Yi2019,Song2022,Uji2023} and superconductor with bulk $T_{c} \approx \SI{7.2}{K}$ \cite{Kuroiwa2007,Rhee2010,Yamada2021,Zhong2024}. The relatively high $T_{c}$, especially given the small density-of-states around $E_{\mathrm{F}}$, has led to speculation that its SC may have an unconventional origin \cite{Yamada2021}. However, this has been contested in other reports and there is a need for more detailed investigation \cite{Kuroiwa2007,Muechler2019}.

\begin{figure*}
\centering
\includegraphics[scale=1]{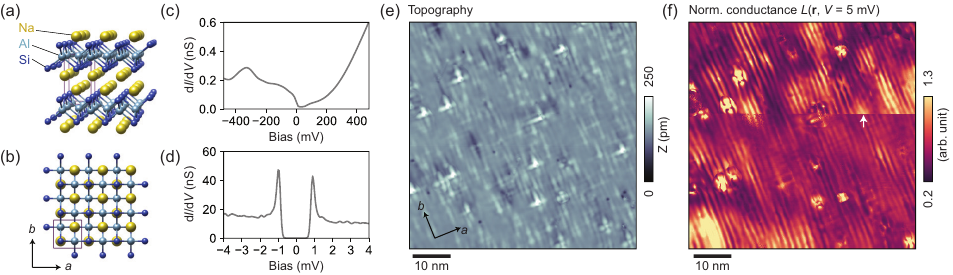}
\caption{\label{fig:1}
Overview of the NaAlSi surface as seen using STM. (a) Depiction of the layered crystal structure and (b) top-down view of its Na-terminated cleaved surface, which has $C_{4v}$ symmetry. Structures are depicted using VESTA \cite{VESTA}. (c) $\frac{\mathrm{d}I}{\mathrm{d}V}(V)$ conductance curve showing overall semimetallic behavior with a `v'-shaped minimum slightly above $E_{\mathrm{F}}$ ($V_{\mathrm{set}} = \SI{480}{mV}$, $I_{\mathrm{set}} = \SI{100}{pA}$, $V_{\mathrm{mod}} = \SI{10}{mV}$). (d) Conductance curve acquired around $E_{\mathrm{F}}$, with effective temperature $T_{\mathrm{eff}} \approx  \SI{350}{mK}$, showing the superconducting gap ($V_{\mathrm{set}} = \SI{50}{mV}$, $I_{\mathrm{set}} = \SI{250}{pA}$, $V_{\mathrm{mod}} = \SI{50}{\mu V}$). (e) Typical STM topograph of the Na-terminated surface ($V_{\mathrm{set}} = \SI{30}{mV}$, $I_{\mathrm{set}} = \SI{100}{pA}$). The lattice vector orientations are marked with black arrows. (f) Image extracted at $V = \SI{5}{mV}$ from simultaneously acquired, normalized conductance ($V_{\mathrm{mod}} = \SI{0.5}{mV}$). 
}
\end{figure*}

The structure of NaAlSi, depicted in Fig. 1(a), belongs to the space group \textit{P}4/\textit{nmm}, and cleavage results in flat Na-terminated surfaces with $C_{4v}$ symmetry as shown in Fig. 1(b). Crystals of NaAlSi were synthesized and measured using STM as described in Appendix A. Tunneling conductance spectroscopy measured at the surface over a wide energy range, shown in Fig. 1(c), yields a curve broadly consistent with a semimetal. Figure 1(d) shows spectroscopy in a narrow range around $E_{\mathrm{F}}$ and at effective electron temperature $T_{\mathrm{eff}} \approx \SI{350}{mK}$. This shows fully gapped SC with $\Delta \approx \SI{1}{meV}$.

Figure 1(e) shows a typical STM topograph of the Na-terminated surface. Although the atomic lattice cannot readily be observed, the lattice orientation can be established from faint Bragg peaks in the fast Fourier transform (FFT) of a conductance image (see below). To image the local density-of-states (LDOS) we first acquire tunneling conductance $\frac{\mathrm{d}I}{\mathrm{d}V}(\mathbf{r},V)$ and, to mitigate artifacts of STM tip height variations, compute the normalized conductance $L(\mathbf{r}, V) = [\frac{\mathrm{d}I}{\mathrm{d}V}(\mathbf{r},V)]/[I(\mathbf{r}, V)/V]$ \cite{Kohsaka2007,Macdonald2016}. Figure 1(f) shows the image $L(\mathbf{r}, V = \SI{5}{mV})$ in which we see charge stripe order with strong spatial fluctuations both in amplitude and in phase. The charge order seems to be weakly pinned: The stripe configuration can undergo abrupt local reconfigurations over time, under ordinary scanning conditions. Such a reconfiguration is captured at the horizontal line marked with a white arrow. This fragility against external perturbations suggests that the order is likely to be purely electronic in origin.

\begin{figure}
\centering
\includegraphics[scale=1]{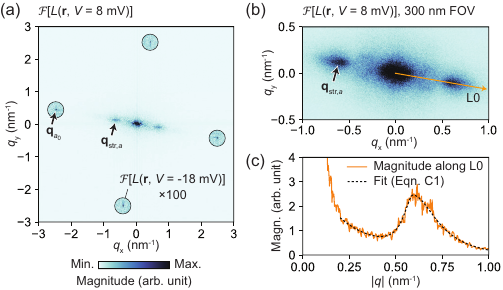}
\caption{\label{fig:2}
Characterization of charge stripe order in $q$. (a) Fast Fourier transform (FFT) of $L(\mathbf{r}, V)$, written as $\mathcal{F} \left[ L(\mathbf{r}, V) \right]$ ($V_{\mathrm{set}} = \SI{30}{mV}$, $I_{\mathrm{set}} = \SI{100}{pA}$, $V_{\mathrm{mod}} = \SI{0.5}{mV}$). The Bragg peaks, at $\mathbf{q}_{a_{0}}$, are not visible simultaneously with the stripes. They become visible only at particular energies and after enhancement of their intensity [100-fold in panel (a)]. The pair of lobes, one of which is marked as $\mathbf{q}_{\mathrm{str}, a}$, describe the charge stripes. (The subscript $a$ denotes a direction parallel with the Bragg vector $\mathbf{q}_{a_{0}}$.) (b) A high-resolution view of the low-$q$ region. [The source $L(\mathbf{r})$ images for both (a) and (b) are shown in Appendix B.] (c) The magnitude of $\mathcal{F} \left[ L(\mathbf{r}, V=\SI{8}{mV}) \right]$ sampled along linecut L0 in (b). The subsequent fitting procedure is described in Appendix C. 
}
\end{figure}

We describe the charge stripe order more quantitatively by analyzing FFTs of $L(\mathbf{r}, V)$, denoted $\mathcal{F} \left[ L(\mathbf{r}, V) \right]$. Figure 2(a) shows this for another cleaved surface. The Bragg vectors, and the wavevector that corresponds to the stripe order, are labeled as $\mathbf{q}_{a_{0}}$ and $\mathbf{q}_{\mathrm{str}, a}$ respectively. The lobes of intensity around $\mathbf{q}_{\mathrm{str}, a}$ have considerable width, and to better characterize them we obtain higher $q$ resolution for another FFT by imaging $L(\mathbf{r}, V=\SI{8}{mV})$ over a $300 \times \SI{300}{nm^{2}}$ field-of-view. The resulting $\mathcal{F} \left[ L(\mathbf{r}, V) \right]$ image is shown in Fig. 2(b). Using Lorentzian fitting (see Appendix C) to a linecut through $\mathcal{F} \left[ L(\mathbf{r}, V) \right]$ parallel to $\mathbf{q}_{a_{0}}$, we estimate the central $q$ vector of the aforementioned lobes, i.e. $\mathbf{q}_{\mathrm{str}, a}$. This gives $\mathbf{q}_{\mathrm{str}, a} \approx 0.235 \mathbf{q}_{a_{0}} $, meaning that the period in the stripe-perpendicular direction is around $4.25 a_{0}$ and the stripes are incommensurate with the lattice. (An observation of the detailed registry between charge stripe and atomic modulations is shown in Appendix D.) The same fitting procedure performed on similar data acquired on a second sample resulted in $\mathbf{q}_{\mathrm{str}, a} \approx 0.25 \mathbf{q}_{a_{0}} $, so that the stripe periodicity was then around $4 a_{0}$.

\begin{figure}
\centering
\includegraphics[scale=1]{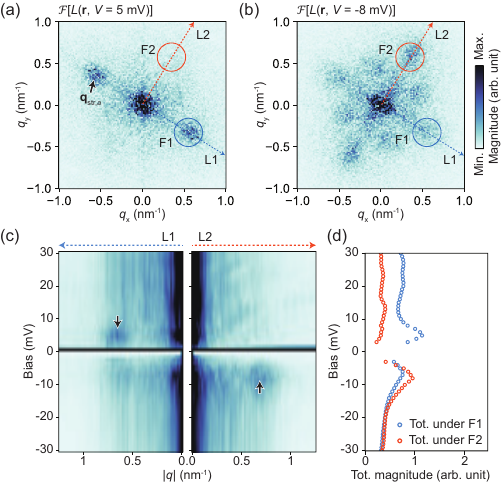}
\caption{\label{fig:3}
Energy dependence of stripe intensity. (a) and (b) $\mathcal{F} \left[ L(\mathbf{r}) \right]$ images extracted above and below $E_{\mathrm{F}}$, respectively ($V_{\mathrm{set}} = \SI{30}{mV}$, $I_{\mathrm{set}} = \SI{100}{pA}$, $V_{\mathrm{mod}} = \SI{0.5}{mV}$). (Source $L(\mathbf{r})$ images are shown in Appendix B.) (c) Linecuts through $\mathcal{F} \left[ L(\mathbf{r}, V) \right]$ taken along the lines L1 and L2 marked in (a). Black arrows mark the features characterizing charge order. (d) The energy-dependence of total intensities under filters F1 and F2, located around the vector $\mathbf{q}_{\mathrm{str},a}$ obtained by 2D Gaussian fitting to the image in (a), and around its 90$^{\circ}$-rotated partner $\mathbf{q}_{\mathrm{str},b}$, respectively. The values near $E_{\mathrm{F}}$ are removed due to severe noise resulting from the vanishing denominator in the normalization. 
}
\end{figure}

An interesting feature of the charge order is seen when we examine the energy dependence of the pattern in $\mathcal{F} \left[ L(\mathbf{r}, V) \right]$, paying attention to the relative intensities near $\mathbf{q}_{\mathrm{str}, a}$ and its 90$^{\circ}$-rotated partner, $\mathbf{q}_{\mathrm{str}, b}$. Figures 3(a) and 3(b) show $\mathcal{F} \left[ L(\mathbf{r}, V) \right]$ images extracted above and below $E_{\mathrm{F}}$. The image above $E_{\mathrm{F}}$ shows intensity at $\mathbf{q}_{\mathrm{str}, a}$ as described above. The image below $E_{\mathrm{F}}$ shows the coexistence of intensity around both $\mathbf{q}_{\mathrm{str}, a}$ and $\mathbf{q}_{\mathrm{str}, b}$. Figure 3(c) shows linecuts through $\mathcal{F} \left[ L(\mathbf{r}, V) \right]$, taken parallel to $\mathbf{q}_{\mathrm{str}, a}$ and $\mathbf{q}_{\mathrm{str}, b}$. The energy-dependence curves for the total intensities around $\mathbf{q}_{\mathrm{str}, a}$ and $\mathbf{q}_{\mathrm{str}, b}$ are shown in Fig. 3(d). The stripe pattern of each orientation shows a distinct peak, one above and one below $E_{\mathrm{F}}$, and with their intensity maxima separated by an energy of $\sim \SI{13}{meV}$.

\begin{figure}
\centering
\includegraphics[scale=1]{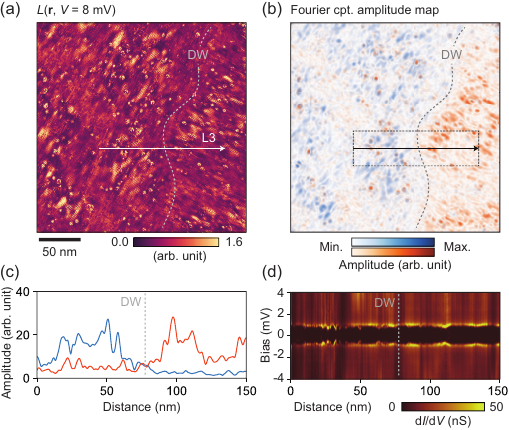}
\caption{\label{fig:4}
Observation of domains of charge stripe order. (a) Large field-of-view $L$ image exhibiting two domains of stripe order with differing orientations ($V_{\mathrm{set}} = \SI{30}{mV}$, $I_{\mathrm{set}} = \SI{100}{pA}$, $V_{\mathrm{mod}} = \SI{0.5}{mV}$). Fourier filtering is used to obtain amplitude maps for the Fourier components using filters similar to those in Fig. 3. The resulting maps are superimposed with separate color scales in (b). (c) Profiles through both amplitude maps [averaged over the short axis of the dashed rectangle in (b)], and (d) conductance measurement of the SC gap along the path L3 marked with a solid arrow ($V_{\mathrm{set}} = \SI{50}{mV}$, $I_{\mathrm{set}} = \SI{250}{pA}$, $V_{\mathrm{mod}} = \SI{50}{\mu V}$, $T_{\mathrm{eff}} \approx  \SI{350}{mK}$). The approximate DW location is marked with a gray dashed line.
}
\end{figure}

Figure 4(a) shows a large field-of-view $L$ map featuring a domain wall (DW) between charge stripe domains that differ in orientation by 90$^{\circ}$. The domains can be highlighted by obtaining Fourier amplitude maps extracted by filtering the FFT. This is done for the components around $\mathbf{q}_{\mathrm{str}, a}$ and $\mathbf{q}_{\mathrm{str}, b}$, and the resulting amplitude maps are overlaid together in Fig. 4(b). A linecut through these amplitude maps and $\frac{\mathrm{d}I}{\mathrm{d}V}(V)$ spectra acquired approximately along the same path are shown in Fig. 4(c) and (d). We see no obvious change in the SC gap size at the location of the DW, which by itself suggests neither a competitive nor cooperative relationship between charge stripe and SC orders. Most importantly, the existence of domains separated by a narrow DW suggests either spontaneous symmetry-breaking order if background strain is negligible or, at least, a high smectic susceptibility such that the stripe order shows a dramatic response to even a weakly varying strain field \cite{Guo2024,Butler2025}.

\begin{figure}
\centering
\includegraphics[scale=1]{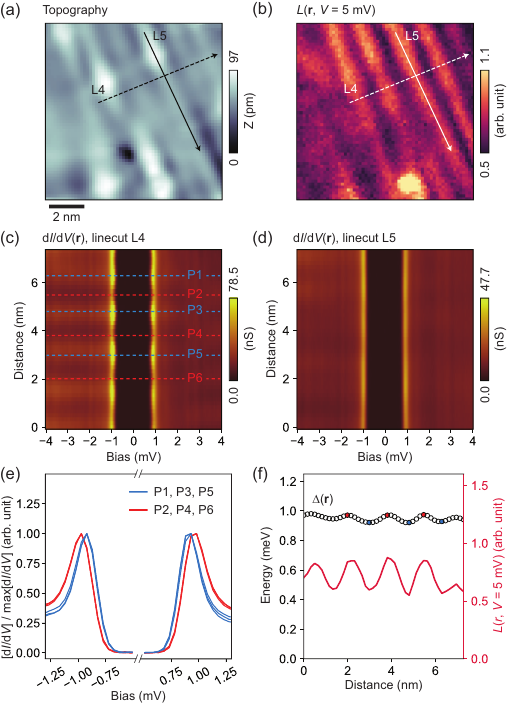}
\caption{\label{fig:5}
Microscopic correspondence between charge stripe and SC orders. (a) A topograph ($V_{\mathrm{set}} = \SI{50}{mV}$, $I = \SI{250}{pA}$) and (b) simultaneous $L$ map in a small field of view. (c) and (d) Linecuts extracted from the simultaneous $\frac{\mathrm{d}I}{\mathrm{d}V}(V)$ conductance ($V_{\mathrm{mod}} = \SI{50}{\mu V}$, $T_{\mathrm{eff}} \approx  \SI{350}{mK}$), along paths L4 and L5 marked in both (a) and (b) with dashed and solid lines, respectively. (e) $\frac{\mathrm{d}I}{\mathrm{d}V}(V)$ curves sampled at the peaks and troughs of the modulation in (c). (f) The value of the local SC order parameter $\Delta(\mathbf{r}) = \left[ \left| \Delta_{\mathrm{UCP}}(\mathbf{r}) \right| +  \left| \Delta_{\mathrm{LCP}}(\mathbf{r}) \right| \right] /2$ (black circles) obtained using the fitting procedure described in Appendix E, and the corresponding variation of $L$ along path L4. The modulations of $\Delta(\mathbf{r})$ and $L(\mathbf{r})$ are in phase.
}
\end{figure}

We now examine the microscopic relationship between SC and charge stripe orders on the scale of the stripe period. Figures 5(a) and 5(b) show topographic and $L$ images, respectively, in a small field-of-view. After measuring $\frac{\mathrm{d}I}{\mathrm{d}V}(\mathbf{r}, V)$ throughout this field-of-view, we extract linecuts perpendicular and parallel to the stripes [along the paths labeled in Fig. 5(a) as L4 and L5, respectively], and these are shown in Fig. 5(c) and 5(d). The SC coherence peaks show a subtle spatial modulation along the path perpendicular to the stripes, but are nearly uniform on the path along a single stripe. Figure 5(e) shows selected $\frac{\mathrm{d}I}{\mathrm{d}V}(V)$ curves at the peaks and troughs of the modulation seen in Fig. 5(c), with locations indicated as blue and red dashed lines. The overall intensity of each $\frac{\mathrm{d}I}{\mathrm{d}V}(V)$ curve is location-dependent, and this dependence is affected by the varying STM tip height. Therefore, for ease of comparison the curves are each normalized according to the maximum conductance. (The segments of the curves at positive and negative bias are treated separately.) Figure 5(e) then shows a clear variation of the energy spacing between the upper and lower SC coherence peaks, namely $2 \Delta$. In Fig. 5(f) we show the result of a phenomenological fit used to estimate and resolve the variation of $\Delta(\mathbf{r})$ (see Appendix E for details). The resulting variation of $\Delta(\mathbf{r})$ is shown in Fig. 5(f) alongside the value of $L(\mathbf{r}, V=\SI{5}{mV})$ along the same path. 

The modulation of the SC gap resembles some claimed observations of Cooper pair density wave phenomena \cite{Liu2021,Chen2021}. In systems where a prior CDW exists, the SC order parameter $\Delta$ is expected to have a spatially modulated component $\Delta(\mathbf{r})$ that follows the charge density modulation, having the same characteristic wavevector(s) \cite{Liu2021,Machida1981}. In that case the CDW can be said to be the primary order while the pair density wave is a subsidiary or secondary order. In the present case the primary order is not a CDW, but is instead liquid crystal-like, and so instead of a secondary pair density wave, the modulation of SC order here might be termed a secondary `pair liquid crystal'. In Fig. 5(f), $L$ can be taken as a proxy for the local charge density, and it is then natural that the variation of $\Delta(\mathbf{r})$ follows that of $L(\mathbf{r})$.

\begin{figure}
\centering
\includegraphics[scale=1]{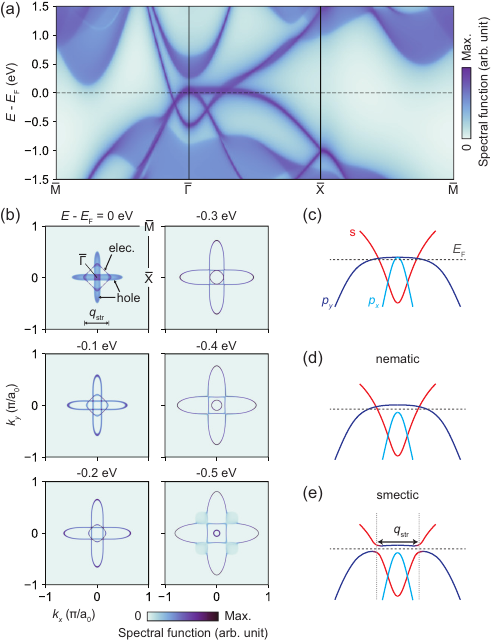}
\caption{\label{fig:6}
Calculated surface spectral function for NaAlSi. (a) Spectral function plotted along the high-symmetry lines of the Brillouin zone. (b) Constant-energy cuts through the spectral function at selected, successively decreasing energies. The length of $\mathbf{q}_{\mathrm{str}}$ is indicated in the first panel. (c)--(e) Schematics of the band structure na\"{i}vely predicted using DFT, the band structure after lifting of the degeneracy between the $p_{x}$- and $p_{y}$-derived hole pockets, and after formation of stripe order.
}
\end{figure}

To begin to understand the origin of the observed charge order, we calculate the surface electronic band structure using the density functional theory (DFT) framework as described in Appendix F. Figure 6(a) shows the calculated spectral function along high-symmetry lines of the surface Brillouin zone. The overall features of the surface bands are found not to differ substantially from those previously calculated for the bulk, because the bulk bands have fairly weak dispersion along the $k_{z}$ axis \cite{Rhee2010,Muechler2019}. The salient features are a $\overline{\Gamma}$-centered electron-like pocket of Al $s$-orbital origin and, also around $\overline{\Gamma}$, two large oblong hole pockets of Si $p$-orbital origin \cite{Rhee2010,Yi2019,Muechler2019}, which have orthogonal orientations and each extend along one of the $\overline{\mathrm{\Gamma X}}$ axes. Notably, the hole pockets only slightly cross the Fermi level from below, and do so with a remarkably flat dispersion over a large range of $k$. Figure 6(b) shows selected constant-energy contours of the spectral function at $E_{\mathrm{F}}$ and at successively lower energies. 

Lowering of rotational symmetry from $C_{4v}$ to $C_{2v}$ can be explained by a lifting of degeneracy between the $p_{x}$ and $p_{y}$ hole pockets, which are depicted schematically in Fig. 6(c), due to the band Jahn-Teller or an equivalent effect. This may be energetically favored due to the large energy saving available if the flat top of one hole pocket sinks below $E_{\mathrm{F}}$. This would lead to a state that breaks the crystal's rotational symmetry but retains its translational symmetry, as depicted in Fig. 6(d), namely a nematic state. Breaking of translational symmetry additionally requires that the band structure undergoes a reconstruction \textit{via} a nesting-driven instability with the wavevector $\mathbf{q}_{\mathrm{str}}$, and with this in mind we indicate its length in the first panel of Fig. 6(b). This shows that $\mathbf{q}_{\mathrm{str}}$ closely matches the wavevector connecting regions along either $\overline{\mathrm{X \Gamma X}}$ axis where the electron band crosses through the flat top of one of the hole bands. The appearance of a new periodicity with wavevector $\mathbf{q}_{\mathrm{str}}$ allows an effective shrinking of the Brillouin zone such that a gap can open at the new zone boundary. This then allows a gap to open where the electron and hole bands cross, as depicted in Fig. 6(e), leading to a further energy saving, and resulting in the observed smectic behavior. (This does not lead to an insulator because the bands cross $E_{\mathrm{F}}$ elsewhere in the Brillouin zone.) This scenario naturally explains why a clearer $C_{2v}$ pattern is seen above $E_{\mathrm{F}}$ but a pattern with symmetry closer to $C_{4v}$ is observed below [recall the pattern in Fig. 3(b), as compared to that in 3(a)]. Only one of the hole bands interacts with the electron band above $E_{\mathrm{F}}$, resulting in stripes, while below $E_{\mathrm{F}}$ both hole bands do so.


In summary, we have described charge stripe order that breaks both rotational and translational symmetries of the host lattice. It is short-ranged, having a coherence length only a few times the modulation period (Appendix C), and its spatial fluctuations appear to be re-configurable under mild electronic perturbation, as seen in Fig. 1(f). 
The observed phenomena are best described as a smectic liquid crystal-like behavior. Although the canonical phenomenologies for a smectic phase and a CDW partially overlap, the present observations would have to be interpreted as a very unusual CDW: It breaks rotational symmetry (unusual for CDWs unless accompanied by orthorhombicity \cite{Que2025}), it has exceptionally strong spatial fluctuations and, especially, its two orthogonal symmetry-breaking LDOS configurations reside at different energies.
The smectic and SC orders are found to be intertwined such that the SC gap $\Delta$ is modulated in phase with the striped LDOS, leading to an unusual secondary `Cooper pair liquid crystal'. Finally, we propose that the driving mechanism may involve a band Jahn-Teller-like lifting of degeneracy between $p_{x}$ and $p_{y}$ bands, incentivized by the large energy saving available due to the bands' flat dispersion in proximity to $E_{\mathrm{F}}$. This is followed by a nesting-driven instability where the remaining Fermi contour is crossed by the coexisting $s$ band. This represents a rare case of electronic liquid crystal behavior in a $p$-orbital material, and extends investigations of the relationship between superconductivity and electronic liquid crystal behavior beyond the well-known $d$ orbital examples.

\section*{Acknowledgements}
We are grateful to  T. Machida and M. Naritsuka for assistance, and to T. Nakamura, Y. Fujisawa, Y. Okada, S.-Y. Guan and T.-M. Chuang for helpful discussions. This work was supported by JSPS KAKENHI Grants No. JP20H02820, No. JP23H04860, No. JP25K21692, and No. JP25H01252, and also by a Grant-in-Aid for Scientific Research on Innovative Areas `Quantum Liquid Crystals' (KAKENHI Grant No. JP19H05824 and No. JP19H05825). M.-C. J. acknowledges support from RIKEN's IPA program.

\section*{Data availability}
The data that support the findings presented here are available from the corresponding authors upon reasonable request.

\appendix

\section{Crystal synthesis and STM measurements}

Crystals were synthesized as described previously \cite{Yamada2021}. They were glued to sample plates, and prepared for later \textit{in situ} cleavage, in a N$_{2}$ glovebox due to the rapid degradation upon exposure to air. They were then loaded into an ultra-high vacuum chamber ($P \sim \SI{e-10}{Torr}$) before cleaving at about $\SI{77}{K}$, after which they were quickly inserted into a modified Unisoku USM1300 low-temperature STM system \cite{Hanaguri2006}. STM measurements were performed using electro-chemically etched tungsten tips that were characterized and conditioned using field ion microscopy followed by repeated mild indentation at a clean Cu(111) surface. Measurements were performed at $T = \SI{1.5}{K}$ unless otherwise stated in the relevant text. Tunneling conductance was measured using the lock-in technique with bias modulation of frequency $f_{\textrm{mod}} = \SI{617.3}{Hz}$ and an amplitude $V_{\mathrm{mod}}$ specified in the caption describing each measurement. The effective electron temperature $T_{\mathrm{eff}}$ was estimated by fitting a Dynes curve to a $\frac{\mathrm{d}I}{\mathrm{d}V}(V)$ spectrum measured using a superconducting Al tip on Cu(111).

Topography maps, conductance maps and their Fourier transforms are plotted using perceptually uniform colormaps \cite{Thyng2016}.

\section{Source $L(\mathbf{r})$ images for $\mathcal{F} \left[ L(\mathbf{r}) \right]$ images displayed in Figs. 2 and 3}

Figure 7. shows the $L(\mathbf{r})$ images used to obtain the Fourier transform results displayed in Figs. 2 and 3. Figures 7(a) and 7(b) show the small and large field-of-view data used to obtain the overview of $q$ space in Fig. 2(a), and the high-resolution view of the low-$q$ region in Fig. 2(b). Figures 7(c) and 7(d), acquired on a different sample, show the $L(\mathbf{r})$ images extracted for positive and negative energies in the same field-of-view, whose $\mathcal{F} \left[ L(\mathbf{r}) \right]$ images are shown in Fig. 3(a) and 3(b) above. As is indicated above for the $\mathcal{F} \left[ L(\mathbf{r}) \right]$ image, it is also clear from the $L(\mathbf{r})$ images that the LDOS adopts a distribution that is more consistent with $C_{4v}$ symmetry as a negative energy.

\begin{figure}[b]
\centering
\includegraphics[scale=1]{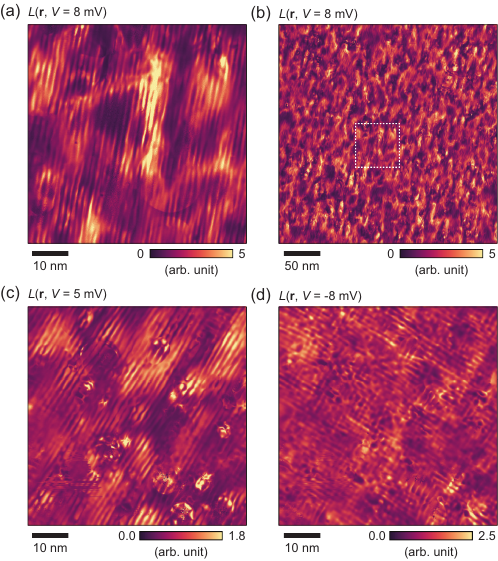}
\caption{\label{fig:7}
Source $L(\mathbf{r})$ images used to obtain the Fourier transform results displayed in Figs. 2 and 3. (a) and (b) Images used to obtain $\mathcal{F} \left[ L(\mathbf{r}) \right]$ in Figs. 2(a) and 2(b), respectively. The smaller field-of-view is marked in the larger one with a white dashed line. (c) and (d) Images used to obtain those in Figs. 3(a) and 3(b), respectively.}
\end{figure}

\section{Lorentzian fitting for charge stripe order parameter}

Fitting to the Fourier magnitude sampled along linecut L0 and shown in Fig. 2(c) was performed using a skewed Lorentzian lineshape \cite{Naritsuka2025} atop an exponentially decaying background:

\begin{equation}
I(q) = \frac{A}{\pi} \left[ \frac{1}{ \gamma \left[ 1 + \frac{ \left| q - q_{\mathrm{str}}\right|^{2}}{ \gamma^{2} \left( 1 + \mathrm{sgn} (q - q_{\mathrm{str}}) \right) } \right] } \right] + e^{-\lambda q} + c.
\end{equation}

Fitting to the data sampled along a linecut passing through $\mathbf{q}_{\mathrm{str}, a}$, but perpendicular to it, was performed using a simple Lorentzian lineshape:

\begin{equation}
I(q) = \frac{A}{\pi} \left[ \frac{\gamma}{(q - q_{\mathrm{str}})^{2} + \gamma^{2}} \right] + c.
\end{equation}

The constant $c$ accounts for the uniform noise floor in $q$ space. In each case $\gamma$ describes the fluctuations of the order and a coherence length can be obtained as $\xi = 1/ \gamma$. Because the charge stripe order breaks rotational symmetry, two coherence lengths, perpendicular and parallel to the stripe orientation, are needed for a full description. These are extracted from the linecut L0 and the perpendicular cut through $\mathbf{q}_{\mathrm{str}, a}$ respectively. The fitted functions give the values $\xi_{\perp} \approx \SI{9.07}{nm}$ and $\xi_{\parallel} \approx \SI{20.33}{nm}$.

\section{Microscopic registry of atomic and charge stripe modulations}

\begin{figure}
\centering
\includegraphics[scale=1]{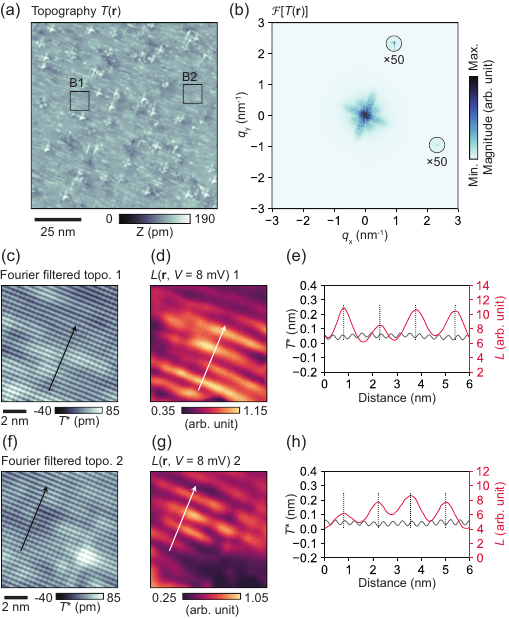}
\caption{\label{fig:8}
Microscopic registry of atomic and charge stripe modulations. (a) Topography ($V = \SI{30}{mV}$, $I_{\mathrm{set}} = \SI{100}{pA}$) and (b) its FFT, with the signals around the locations of the Bragg vectors enhanced for clarity. (c) and (d) Zoom-in views of topography in boxes B1 and B2 marked in (a), and with the atomic corrugations enhanced 500-fold using inverse Fourier filtering, the result of which we call $T^{\ast}(\mathbf{r})$. (e) and (f) Simultaneously measured conductance $L$ (using $V_{\mathrm{mod}} = \SI{0.5}{mV}$) in the same fields-of-view. (g) and (h) Linecuts through  $T^{\ast}(\mathbf{r})$ and  $L(\mathbf{r})$ along the paths marked in (c) and (d), respectively.}
\end{figure}

In Fig. 2 above we observe from the FFTs of $L$ images that the periodicity of the stripe order is incommensurate with the atomic lattice. Here we show the microscopic correspondence in $r$-space between the stripe modulations and the lattice. Because the atomic lattice modulations are not readily observable in topography images, we enhanced them using an inverse Fourier transform technique. Figure 8(a) shows a topography image and Fig. 8(b) shows its FFT with the signals around each of the Bragg vectors enhanced 50-fold for clarity. We filter for these signals, and multiple them by a factor of 500 before inverse transforming them and adding them back to the original topography. We then inspect two small fields-of-view, in boxes labeled B1 and B2 in Fig. 8(a). We compare the manipulated topography maps $T^{\ast}(\mathbf{r})$ shown in Fig. 8(c) and 8(d), with the simultaneously acquired conductance maps $L(\mathbf{r}, V=\SI{8}{mV})$ shown in Fig. 8(e) and 8(f). Figures 8(g) and 8(h) shown linecuts through both $T^{\ast}(\mathbf{r})$ and $L(\mathbf{r})$ images, with the local maxima of $L$ marked by dashed lines. These comparisons show that there is no strict registry between the charge stripe and atomic corrugations, and that they are microscopically incommensurate.

\section{SC coherence peak fitting}

In Fig. 5(f) we show results of fitting to the upper and lower coherence peaks seen in $\frac{\mathrm{d}I}{\mathrm{d}V}(V)$ curves. The coherence peaks appear to be qualitatively unlike those of the Dynes curve expected in the BCS paradigm. (The tail of each coherence peak falls off too quickly.) Because of this we resort to using a phenomenologically motivated fitting function. We use separate functions for the upper and lower coherence peaks (UCP and LCP), each a Lorentzian lineshape atop a hypertangent background, written as

\begin{equation}
\begin{aligned}
\rho(V) & = \frac{A_{_{\mathrm{UCP}}}}{\pi} \left[ \frac{\gamma_{_{\mathrm{UCP}}}}{(eV - \Delta_{_{\mathrm{UCP}}})^2 + \gamma_{_{\mathrm{UCP}}}^2} \right] \\
& + B_{_{\mathrm{UCP}}} \mathrm{tanh} \left(w_{_{\mathrm{UCP}}} \left(eV - E_{_{\mathrm{UCP,\mathrm{tanh}}}} \right) \right) \\
\end{aligned}
\end{equation}

for the UCP, and similarly but with a reversed energy scale for the LCP. The assessed value of the gap parameter is then $\Delta(r) = \left[ \left| \Delta_{_{\mathrm{UCP}}}(r) \right| + \left| \Delta_{_{\mathrm{LCP}}}(r) \right| \right] /2$.

\section{DFT calculations}

The calculation of the electronic structure of NaAlSi was performed using DFT as implemented in the Vienna \textit{ab initio} simulation package (VASP) \cite{Kresse1996a,Kresse1996b}.
All the calculations were performed with the projector-augmented wave (PAW) \cite{Blochl1994} pseudopotential and the generalized gradient approximation (GGA) in the form of Perdew-Burke-Ernzerhof (PBE) \cite{Perdew1996,Kresse1999}.
The plane-wave cutoff energy was $\SI{450}{eV}$ and a $\Gamma$-centered $12 \times 12 \times 12$ $k$-mesh was used to describe the electronic structure. The experimental lattice parameters and atomic positions for NaAlSi are used, with values of $a_{0} = \SI{4.125}{\AA}$ and $c_{0} = \SI{7.374}{\AA}$ \cite{Yamada2021}. Based on the DFT electronic band structure, we construct the atomic Wannier functions using 16 orbitals including Al $s$, $p$ and Si $s$, $p$ \cite{Pizzi2020}. For the surface spectral function, we utilize the iterative Green's function \cite{LopezSancho1985} to compute the (001) surface bands of NaAlSi by constructing a semi-infinite slab using the Wannier tight-binding Hamiltonian as implemented in the WannierTools package \cite{Wu2018}. We also perform a DFT claculation of an 11-layer NaAlSi slab with Na terminations to find good agreement in the surface band results between the two methods.

\end{document}